\documentstyle[prl,aps,epsf,floats]{revtex}

\begin{document}
\draft

\twocolumn[\hsize\textwidth\columnwidth\hsize\csname%
@twocolumnfalse\endcsname

\title{Effect of reheating on electroweak baryogenesis}

\author{Ariel M\'{e}gevand\cite{email}}
\address{
Centro At\'omico Bariloche and Instituto Balseiro,
Comisi\'on Nacional de Energ\'{\i}a At\'omica and
Universidad de Cuyo,\\ 8400 S. C. de Bariloche, Argentina
}

\date{\today}

\maketitle

\begin{abstract}
The latent heat released during the expansion of bubbles in
the electroweak phase transition reheats the plasma and
causes the bubble growth to slow down.
This decrease of the bubble wall velocity affects the result of
electroweak baryogenesis. Since the efficiency of
baryogenesis peaks for a wall velocity  $\sim 10^{-2}$, the resulting
baryon asymmetry can either be enhanced or suppressed,
depending on the initial value of the wall velocity.
We calculate the evolution of the
phase transition taking into account the release of latent heat.
We find that,
although in the SM the baryon production is enhanced by this effect, in the
MSSM it causes a suppression to the final baryon
asymmetry.  \end{abstract}


]

Generating the baryon asymmetry of the universe (BAU) at the
electroweak phase transition is a very attractive idea
\cite{reviews}. Although in the minimal Standard Model (SM) it
fails to explain the observed baryon abundance, such a proposal can be
successful in an
extension of the model. It has been shown, for instance, that electroweak
baryogenesis is quantitatively possible in the Minimal Supersymmetric
Standard Model (MSSM), provided that the Higgs boson and the lightest stop
are sufficiently light \cite{bjls97,cqw98}.

According to the standard mechanism \cite{ckn91}, baryogenesis occurs near
the walls of  expanding bubbles that form during the phase transition.
After a bubble is nucleated and begins to grow, its wall quickly reaches a
terminal velocity in the hot plasma \cite{lmt92,t92,dlhll92}. Due to $CP$
violating
interaction of particles with bubble walls, different densities of
left-handed quarks and their antiparticles are built up in front of the
walls. This left-handed asymmetry biases the anomalous baryon number
violating sphaleron interactions in the symmetric phase. As a consequence, a
net baryon asymmetry is generated in front of the walls and immediately
caught by them. In the broken symmetry region inside the bubbles sphaleron
processes are suppressed, so a subsequent washout of the baryon asymmetry is
avoided.

The generated BAU has a strong dependence on the bubble wall velocity. If
the latter is too large, the left-handed density perturbation will pass
so quickly through a given point in space that sphaleron processes will not
have enough time to produce baryons. Thus the resulting BAU will be small.
On the other hand, for very small velocities thermal equilibrium will be
restored, so the baryon asymmetry will be erased by sphalerons and the BAU
will be small again. Consequently, the baryon production has a maximum at an
intermediate wall velocity. Comparison of the baryon number violation time
scale
with the time of passage of the chiral asymmetry \cite{lmt92,ckn92,h95}
gives a wall velocity $v_{w}\sim 10^{-2}$ for maximum baryon asymmetry.
Recent numerical calculations confirm that the BAU tends to peak for such a
small value of $v_{w}$ \cite{ck00,cmqsw00}.

On the other hand, recent calculations of
the friction of the plasma indicate that the wall velocity can be of
that order of magnitude \cite{m00,js00}. However, these estimates of
$v_w$ do
not take into account the hydrodynamics of the phase transition \cite
{h95,ikkl94}. For such small velocities the only effect of hydrodynamics
is a homogeneous reheating of the plasma during the expansion of bubbles.
As $v_w$ is much less than the speed of sound
in the plasma, the latent heat liberated in the
expansion of a bubble is quickly distributed throughout space by a shock
front that precedes the propagating wall of the bubble.
As a consequence of this uniform reheating the bubble expansion
slows down. It was
shown by Heckler \cite{h95} that the decrease of $v_w$ can dramatically
affect the result of electroweak baryogenesis.
The importance of reheating can be estimated \cite{h95} by comparing the
latent heat $L\equiv
T\left( d\left( \Delta V\right)/dT\right) _{T=T_{c}}$, where $\Delta V\left(
T\right) $ is the free energy difference between the symmetric and broken
symmetry phases, with the energy needed to bring the plasma back to the
critical temperature $T_{c}$ from the temperature $T_{n}$ at which
nucleation of bubbles begins, $\Delta \rho =\left( \pi ^{2}g_{*}/30\right)
\left( T_{c}^{4}-T_{n}^{4}\right) $, where $g_{*}\simeq 107$ is the number
of degrees of freedom of the plasma. In the SM, $L$ is at least one order of
magnitude less than $\Delta \rho $, but in the MSSM the two quantities are
of the same order \cite{ikkl94,yo00}, so the effect of reheating can be
important in that case.

In this paper we compute the evolution of the phase transition including
this effect in a simple model with a one-Higgs effective potential. Using
an analytical
approximation for the dependence of the BAU on $v_{w}$, we evaluate the
effect of the decrease of the bubble wall velocity on baryogenesis.
It is known that in the SM the reheating enhances electroweak baryogenesis
\cite{h95}. On the contrary, since
in the MSSM the wall velocity is initially in the range of maximum BAU,
its decrease will cause a suppression to the baryon asymmetry.

We use an  effective potential  of the form
\cite{dlhll92,ikkl94,l77}
\begin{equation}
V\left( \phi ,T\right) =D\left( T^{2}-T_{0}^{2}\right) \phi ^{2}-ET\phi
^{3}+\frac{\lambda}{4} \phi ^{4}\ ,  \label{veff}
\end{equation}
which possesses a first-order phase transition between the critical
temperature $T_{c}=T_{0}/\sqrt{1-E^{2}/\lambda D}$ and $T_{0}$. When the
universe cools below $T_{c}\simeq 100GeV$, bubbles of the broken
symmetry phase begin to nucleate with a probability per unit volume and time
$\Gamma \sim T^{4}e^{-S_{3}\left( T\right) /T}$, where $S_{3}\left[ \phi
\left( r\right) \right] $ is a three-dimensional instanton action that
coincides with the energy of the nucleated bubble \cite{l77}. The
configuration $\phi \left( r\right) $ of the bubble is obtained by finding
an extremum of $S_{3}$. It can be calculated numerically to obtain the
nucleation rate $\Gamma $ \cite{dlhll92} and the radius $r_{0}$ and wall
width $l_{w}$ of the nucleated bubble \cite{yo00} as functions of
temperature.

For the terminal velocity of the bubble wall in the plasma we use the
formula \cite{js00} (see also \cite{lmt92,t92,h95,yo00})
\begin{equation}
v_{w}\left( T\right) \simeq \frac{20Tl_{w}\left( T\right) \Delta V\left(
T\right) }{\eta v\left( T\right) ^{4}}\ ,
\end{equation}
where $\eta $ is a dimensionless friction coefficient accounting for the
viscosity of the plasma, and $v\left( T\right) $ is the minimum of $V\left(
\phi ,T\right) $. The main effect of the reheating will be to decrease the
free energy difference $\Delta V\left( T\right) $
\cite{notalv} and hence the value of $v_{w}$.
The progress of the phase transition is determined by the fraction of
space that is still in the symmetric phase \cite{gw81},
\begin{equation}
f\left( t\right) =\exp \left\{ -\frac{4\pi }{3}\int_{t_{c}}^{t}\Gamma \left(
T^{\prime }\right) r\left( t^{\prime },t\right) ^{3}dt^{\prime }\right\} \ ,
\label{f}
\end{equation}
where $r\left( t^{\prime },t\right) =r_{0}\left( T^{\prime }\right)
+\int_{t^{\prime }}^{t}v_{w}\left( T^{\prime \prime }\right) dt^{\prime
\prime }$ is the radius at time $t$ of a bubble created at $t^{\prime }$.
The variation of temperature with time is given by \cite{h95}
\begin{equation}
\frac{dT}{dt}=\frac{TV^{\prime }\left( T\right) -V\left( T\right) }{\left(
2\pi ^{2}g_{*}/15\right) T^{3}}\frac{df}{dt}-\left( \frac{8\pi ^{3}g_{*}}{
90M_{Pl}^{2}}\right) ^{1/2}T^{3}\ ,  \label{tt}
\end{equation}
where $M_{Pl}=1.22\times 10^{19}GeV$ is the Plank mass. The first term is
the contribution of reheating. It describes the increase of energy
density of the plasma due to the release of latent heat during the phase
transition. The second term is just $-HT$; it accounts for the decrease of
energy density caused by the expansion of the universe. Before and after
the phase transition $df/dt=0$, and Eq.~(\ref{tt}) gives the well known
relation $t= \xi M_{Pl}/T^{2}$, with $\xi \simeq 0.03$. During the
phase transition, the
coupled integro-differential equations (\ref{f},\ref{tt}) must be solved
numerically.

To evaluate the effect of a changing wall velocity on baryogenesis
we need to know how the baryon production depends on $v_{w}$. As the
bubble wall sweeps through space, it leaves behind a baryon density \cite
{ck00}
\begin{equation}
n_{B}=\frac{9\Gamma _{ws}}{v_{w}}\int_{0}^{\infty }n_{L}\left( x\right)
e^{-c\Gamma _{ws}x/v_{w}}dx\ ,  \label{nb}
\end{equation}
where $\Gamma _{ws}\simeq 20\alpha _{w}^{5}T$ \cite{mr99} is the weak
sphaleron rate, and $n_{L}$ is the net left
handed density in front of the wall. The exponential accounts for sphaleron
relaxation of the baryon asymmetry for small velocities. The coefficient $c$
depends on the squark spectrum and is $\sim 10$.

The left-handed density can be assumed to be of the form $n_{L}\left(
x\right) =Ae^{-v_{w}x/D}$ , where $D\sim 100/T$  is an effective diffusion
constant for the chiral asymmetry \cite{hn96}. The constant $A$ depends on
the $CP$ violating force at the bubble wall that sources the asymmetry.
It is in general proportional to the wall velocity, $A\propto v_{w}$, so
integration of Eq.~(\ref{nb}) yields
\begin{equation}
n_{B}=\frac{C}{v_{w}+c\Gamma _{ws}D/v_{w}}\ ,  \label{nbvw}
\end{equation}
where $C$ does not depend on $v_{w}$. This analytic approximation describes
qualitatively the dependence of $n_{B}$ on $v_{w}$. It has a peak at
$v_{\text{peak}}\equiv \sqrt{c\Gamma _{ws}D}\simeq 0.02$. A calculation of
the coefficient $C$ is out of the scope of this paper, since we will only be
interested in relative values $n_{B}\left( v_{w}\right) /n_{B}\left(
v_{0}\right) $. Note that if the initial wall velocity is $v_{0}\gg
v_{\text{peak}}$, then a decrease of $v_{w}$ will produce an enhancement
of the BAU, since in that case $n_{B}\sim v_{w}^{-1}$. On the contrary, for
small velocities $n_{B}\sim v_{w}$, so a velocity decrease will cause a
suppression of the final BAU.

As $v_{w}$ varies during the phase transition, $n_{B}\left( v_{w}\left(
t\right) \right) $ gives only a {\em local} baryon density, generated in a
volume $dV\left( t\right) =-V_{\text{Total}}\dot{f}\left( t\right) dt$.
The final baryon density is the average over the expansion of
bubbles \cite{nota2} $B= \frac{1}{V_{\text{Total}}}
\int n_{B}\left( t\right) dV$. According to Eq.~(\ref{nbvw}), this
is related to the result obtained with a constant wall velocity $v_{0}$ by $
B=Sn_{B}\left( v_{0}\right) $, where the factor
\begin{equation}
S=-\int \frac{df}{dt}\frac{v_{0}+v_{\text{peak}}^{2}/v_{0}}{v_{w}\left(
t\right) +v_{\text{peak}}^{2}/v_{w}\left( t\right) }dt\ .  \label{s0}
\end{equation}
gives the enhancement or suppression due to the effect of reheating.

We consider three sets of parameters for the electroweak
phase transition.
We take as case A the SM with an unrealistically small value of the Higgs
mass, which allows for a sufficiently strong first-order phase transition.
Therefore, we set the values $D=0.2$ and $E=0.006$ \cite{dlhll92}, and we
choose $\lambda =2E$ in order to fulfill the condition $v\left( T_{c}\right)
/T_{c}\gtrsim 1$, which is required for avoiding the washout of the BAU.
For the friction of the plasma in the SM we assume the rough value $\eta
\sim 1$ \cite{lmt92,t92,dlhll92}.

For the MSSM, the one-Higgs potential
(\ref{veff}) can be used as an approximation
in the case in which only one Higgs boson is light \cite{cqw98}.
In this scenario the parameter $E$ can be at most an order of magnitude
larger than in the SM. Although this limiting situation is hardly achieved
in practice, we can use it to simulate, without departing from reasonable
values of the parameters, an extension of
the SM that is most favorable for baryogenesis.
We thus choose for case B the values $E=0.06$ with $\lambda =2E$, and $D\sim
1$.  In this case we assume a
friction coefficient $\eta \simeq 70$, in accordance with recent
calculations for the MSSM with a light right-handed stop \cite{js00}.

As we use it only to compute the dynamics of the phase transition,
$V\left( \phi ,T\right) $ need not be a real perturbative
effective potential.
Instead, we can
regard the field $\phi $ as an effective order parameter, and use
Eq.~(\ref{veff}) to model the phase transition dynamics \cite {h95,ikkl94}.
In that way the parameters $D$, $E$, and $\lambda $ can be chosen so that
the free energy $V$ has the same thermodynamical properties of the theory we
wish to study. The relevant quantities are the latent heat $L$ defined
above, the surface tension of the bubble wall, $\sigma \equiv \int \left(
d\phi /dx\right) ^{2}dx|_{T_{c}}$, and the correlation length, given by $\xi
^{-2}\equiv \partial ^{2}V/\partial \phi ^{2}|_{v\left( T_{c}\right),
T_{c}}$. In our model these parameters are given by $L/T_{c}^{4}=8D\left(
E/\lambda \right) ^{2}\left( 1-E^{2}/\lambda D\right) $, $\sigma
/T_{c}^{3}=2\sqrt{2} E^{3}/3\lambda ^{5/2}$, and $\xi T_{c}=\sqrt{\lambda
/2}/E$. We thus choose the parameters of case C in accordance with recent
non-perturbative lattice simulations of the MSSM in the light right-handed
stop scenario, which give $ L/T_{c}^{4}\simeq 0.4$, $\sigma /T_{c}^{3}\simeq
0.01$, and $\xi T_{c}\simeq 5$ \cite{lr00}. This case represents a more
physical situation, as it allows electroweak baryogenesis for experimentally
viable values of the Higgs mass.

In Fig.~1 we have plotted the evolution of the phase transition for case A.
\begin{figure}[h]
\centerline{\epsfysize=8.3cm\epsfbox{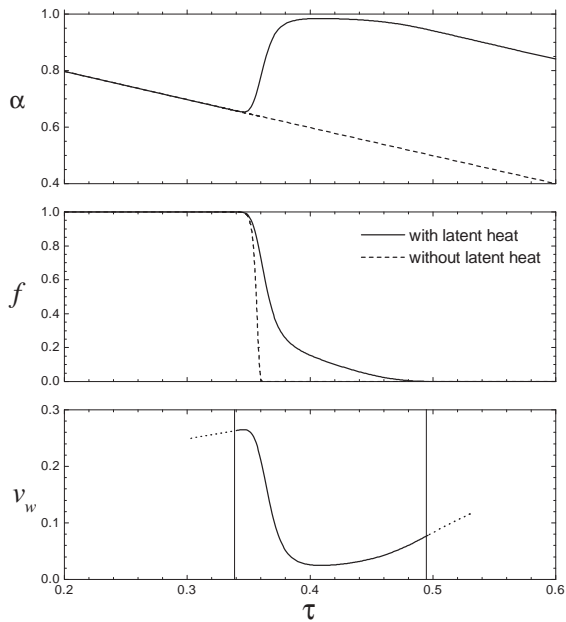}} \caption{Dimensionless
temperature $\alpha $, fraction of volume in the symmetric phase $f$, and
bubble wall velocity $v_{w}$, as functions of dimensionless time $\tau$, for
the parameters of case A, with and without including the release of latent
heat. The vertical lines delimit approximately the phase transition
interval.} \end{figure}
We have defined a dimensionless temperature $\alpha =\left( T-T_{0}\right)
/\left( T_{c}-T_{0}\right) $, and a dimensionless time $\tau =\left(
t-t_{c}\right) /\left( t_{0}-t_{c}\right) $, where
$t_{0}=\xi M_{Pl}/T_{0}^{2}$ is the time at which the universe reaches the
temperature $T_{0}$ if reheating is ignored.
The result obtained when neglecting the reheating is
plotted with dashed lines.
In that case the duration of the phase transition
is so small that all the parameters that enter the BAU can be approximated
by their values at the temperature $T_{n}$ of the onset of nucleation
\cite{yo00}. When the effect of latent heat is included, the plasma heats up
after the beginning of nucleation because the expansion of the universe does
not remove energy fast enough to compensate the release of latent heat. The
bubble wall velocity thus decreases, and the phase transition slows down
until an equilibrium temperature is reached, at which all the released
latent heat goes into expanding the universe. Finally, the phase transition
completes and the temperature decreases again. Due to the reheating the wall
velocity approaches $v_{\text{peak}}$, so the baryon asymmetry results enhanced.
Integration of Eq.~(\ref{s0}) gives $S\simeq 3$.
As expected, the enhancement is not considerable since,
as the latent heat is not very large in this case, $v_{w}$ does not spend a
long time near $ v_{\text{peak}}$.

In case B, in contrast, the latent heat is much larger. The plasma heats
up very close to $T_{c}$ and the temperature remains constant for a
long period of time (see Fig.~2).
\begin{figure}[h]
\centerline{\epsfysize=8.3cm\epsfbox{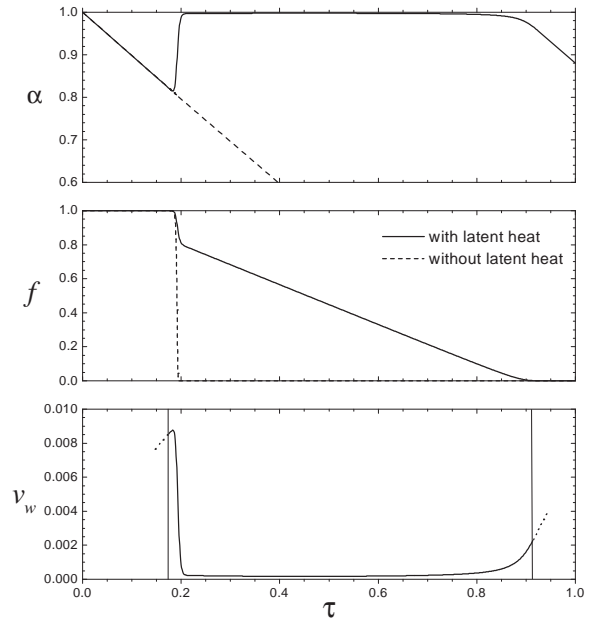}}
\caption{Same as Fig.~1, but for case B.}
\end{figure}
However, the BAU turns out to be suppressed in this case.
Since the initial wall velocity is less in the MSSM due
to the larger friction of the plasma, the decrease of $v_w$
occurs in the region on the left of the peak.
Eq.~(\ref{s0}) gives a suppression factor $S\simeq 0.15$.
Note that the slow growth period, in which $v_{w}\simeq
2\times 10^{-4}\sim 10^{-2}v_0$, gives only a contribution
$\sim 10^{-2}$ to $S$. The
baryon asymmetry is essentially generated in the initial stage in which
bubbles fill a 20\% of the volume of the universe. There exists an
additional suppression with
respect to the maximum BAU, since the initial wall velocity $v_0$ is a
little below $v_{\text{peak}}$. The total suppression factor is $2\left(
v_{0}/v_{\text{peak}}+v_{\text{peak}}/v_{0}\right) ^{-1} S\simeq 0.1$.

The results of case C are plotted in Fig.~3. Here the latent heat is smaller
than in the previous case, so the final velocity is larger, $v_{w}\simeq
5-7\times 10^{-4}$, and the phase transition occurs in a shorter time. As a
consequence, the suppression of the baryon asymmetry due to reheating is
less severe, $S\simeq 0.4$.
\begin{figure}[h]
\centerline{\epsfysize=8.3cm\epsfbox{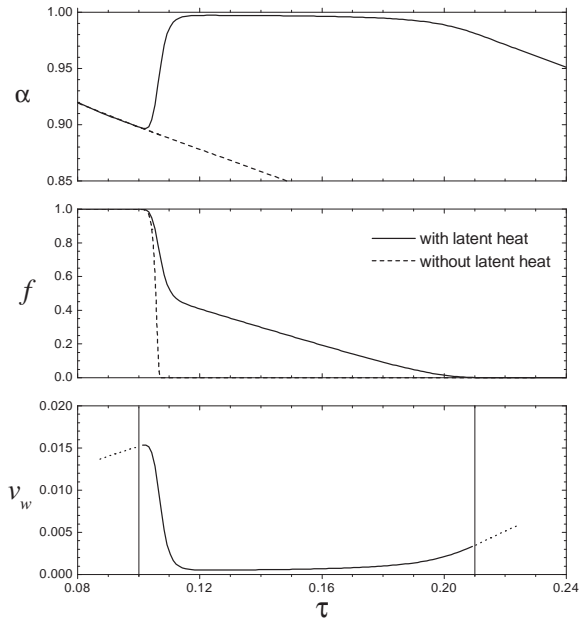}}
\caption{Same as Fig~1, but for case C.}
\end{figure}

Finally, it is important to notice that
in some cases the simplifying approximation
(\ref{nbvw}) may fall short of describing the behavior
of $n_{B}\left( v_{w}\right) $, especially for small $ v_{w}$.
For such cases, further suppression may arise in a more rigorous
calculation. For example, in Ref.~\cite{ck00} a crossing through zero of
$n_B$ occurs at $v_{w}\sim 10^{-3}$. That happens because the $CP$ violating
force changes sign near the bubble wall, and so does the chiral asymmetry.
This gives an opposing contribution to $n_{B}$, which becomes important for
small wall velocities. Hence, in our last two cases baryon number densities
of opposite sign would be generated in different regions of the universe.
The negative contribution of the slow growth period to Eq.~(\ref{s0}) can be
roughly estimated by multiplying the ratio $ n_{B}\left( 10^{-4}\right)
/n_{B}\left( 10^{-2}\right) $ by the fraction of space $\Delta f$ that is
filled during this stage. According to the results of Ref.~\cite{ck00}, that
ratio depends strongly on the bubble wall width and on the gaugino mass
parameters. If we take for instance a value $\sim -1/3$, then in case B,
with $\Delta f\simeq 0.8$, this contribution is of the same order of the
previously generated baryon asymmetry. In case C, with $\Delta f\simeq 0.4$,
the suppression factor would decrease a $30\%$.

To summarize, we have calculated the decrease of the bubble
wall velocity that occurs during the electroweak phase transition as
a consequence of reheating, as well as its effect
on baryogenesis.
In the MSSM this effect tends to suppress the baryon production.
For the light
stop scenario we have found a suppression factor $S\simeq 0.4$. Although
this is
not severe, we must stress that $S$ can be smaller if the
baryon density changes sign for small wall velocities.


\begin{references}
\bibitem{notalv}  $l_{w}$ and $v$ do not vary significantly between $
T_{n}$ and $T_{c}$ \cite{yo00}.
\bibitem{nota2}  It can be seen from our results that the variation of $
v_{w} $ is irrelevant for the diffusion equations that lead to
Eqs.~(\ref{nb}) and (\ref{nbvw}) (see also  \cite {ck00}).
\end{references}
\end{document}